\documentstyle[preprint,aps,epsfig]{revtex}

\newcommand{\etal}{{\it et al.}}	    %%%  et al.

\begin{document}

\preprint{UTF/02-2001-442}

\draft

\title{Coulomb Distortion in Quasielastic $(e,e')$ Scattering on Nuclei: \\
Effective Momentum Approximation and Beyond}
\author{Marco Traini}
\address{
Dipartimento di Fisica, Universit\`a degli Studi di Trento,\\ 
I-38050 POVO (Trento), Italy\\
and Istituto Nazionale di Fisica Nucleare, G.C. Trento}
\maketitle

%\begin{center}
%(\today)
%\end{center}

\begin{abstract}
The role of the effective momentum approximation to disentangle Coulomb distortion
effects in quasielastic $(e,e')$ reactions is investigated. The separation 
of the cross section in longitudinal and transverse
components is discussed including higher order DWBA corrections due 
to the focusing of the electron waves.
The experimental studies performed, in the last few years, 
making use of different approximate treatments are shown to be sometime 
inconsistent. As a consequence some of the longitudinal and transverse 
responses, extracted from the inclusive cross sections cannot be considered 
reliable. A separation procedure based on the effective momentum approximation 
is discussed in connection with the recent experimental data on electron/positron 
quasielastic scattering on $^{12}$C and $^{208}$Pb.
\end{abstract}

\vspace{2mm}

Pacs: 25.30.Fj


$\phantom{Keywords:}$Coulomb corrections.

\newpage

\section{Introduction}
\label{sect0}

The quasielastic electron scattering off nuclei has represented,
in the last 20 years, one of the most successful tools to study
nuclear and nucleon structure properties. Both inclusive $(e,e')$
and exclusive (single arm $(e,e',{\cal N})$ or double arm
$(e,e',{\cal N,N})$) contributed to a deeper understanding of the
many-body structure of strongly interacting systems like light
and heavier nuclei opening the possibility of investigating 
also the {\it in medium} nucleon properties.
In particular the quenching of the longitudinal strength in
inclusive reactions\cite{OrTr91} has been related to partial 
restoration of chiral symmetry in nuclei\cite{brownrhosoy} combined
with effects due to many-body short-range correlations in dense
matter\cite{TrOrLei93}. Similar results have been recently obtained
within a relativistic RPA approach taking into account the {\it in medium} 
modifications of the nucleon structure as described by a quark-meson 
coupling model\cite{saitoetal99}.

However the experimental studies of inclusive and exclusive reactions
induced by electrons have an intrinsic limitation in the case of target 
nuclei with a large number of protons. The strong Coulomb 
field induces a distortion of the wave front which modifies 
the structure of the $(e,e')$ cross section and induces 
sizable effects in the longitudinal/transverse separation
of the electromagnetic responses\cite{uberall,CoHei87,TrTu87,TrTuZg88}.

The theoretical framework to investigate Coulomb corrections 
to the electron-nucleus cross sections is well 
established \cite{uberall} and is called Distorted Wave 
Born Approximation (DWBA) in contrast to the better known
Plane Wave Born Approximation (PWBA) where the incoming and 
outgoing charged leptons are described by (Dirac) plane wave 
neglecting the effect of the Coulomb interaction between the
projectile and the target nucleus.
The application of the DWBA scheme to the quasielastic $(e,e')$ regime is 
in principle straightforward \cite{CoHei87}, even if the numerical 
complications are extremely time consuming. Moreover the 
DWBA cross section cannot be written in a (Rosenbluth) separable 
form to extract charge (longitudinal) and current (transverse) 
responses: a property valid in PWBA only. As a consequence the 
direct numerical application of the DWBA approach cannot help in the 
separation of the structure functions\cite{CoHei87,TrTu87,TrTuZg88}.

The aim of the present work is to demonstrate that an Effective Momentum 
Approximation (EMA) can be defined and used, even in heavier nuclei,  
to disentangle Coulomb corrections
from the experimental cross section once the effective value
of the Colomb interaction between the electron and the nucleus is
experimentally determined. Focusing corrections are automatically 
included at the lowest order of the EMA and higher order corrections
can be estimated both theoretically and experimentally.

In section \ref{sect1} the concept of effective momentum transfer 
is reviewed emphasizing how the Mott cross section can be factorized out
in quasielastic scattering. In section \ref{sect2} higher order
corrections are investigated and the mean value of the Coulomb interaction 
discussed in view of recent experimental results. Numerical
approaches are discussed in section \ref{sect3} and the experimental 
analysis of ($e,e'$) quasielastic data revised 
in section \ref{sect4}. Conclusions are drawn in section \ref{concl}.

\section{The effective momentum transfer}
\label{sect1}

Since the $(e,e')$ DWBA cross section does not assume a 
separable form, longitudinal and transverse components can be extracted, 
in heavy nuclei, only approximately and with the help of theoretical
assumptions. The milestones of this path have been indicated 
by several authors in the past and I would like to follow their
main arguments to demonstrate that the (approximated) PWBA-like form of the DWBA 
cross section must have a structure related to 
a specific physical ingredient: the effective momentum transfer. 
This quantity can be defined only in a phenomenological way because it
is connected to an asymptotic expansion of the cross section, and embodies 
leading corrections to the PWBA cross section, on top of which one has to 
consider higher order effects, if relevant.

\subsection{The eikonal approximation}
\label{sect11}

Czy\.z and Gottfried \cite{CG63}, in their seminal work, 
discussed the break down of the PWBA for heavy nuclei 
and concluded that "{\it one may readily and quite reliably correct for 
this}". They defined the effective momentum transfer in their eq.(2.10)
\begin{equation}
{\bf q}_{\rm eff} = {\bf k}_{\rm i,eff} - {\bf k}_{\rm f,eff} = 
{\bf q} +\left(\hat {\bf k}_{\rm i} - \hat {\bf k}_{\rm f}\right)
{\bar V}_C\,\,,
\label{effmomv}
\end{equation}
where ${\bf q} = {\bf k}_{\rm i} - {\bf k}_{\rm f}$ is the kinematical momentum transfer
as measured in the laboratory frame where the initial (final) electron momentum
${\bf k}_{\rm i}$ (${\bf k}_{\rm f}$) is determined. ${\bar V}_C$ represents the Coulomb
interaction energy between the electron and the target 
nucleus so that its effective momentum in the vicinity of the nucleus
becomes: ${\bf k}_{\rm i,f,eff} = {\bf k}_{\rm i,f} - \hat {\bf k}_{\rm i,f}{\bar V}_C$. 
($E_{\rm i,f} = |{\bf k}_{\rm i,f}|$ are the energies of the incoming and outgoing electrons
as measured in the lab frame and whose masses are neglected in the high-energy limit).

The DWBA cross section as calculated in ref.\cite{CG63} (eq.(2.11))
is found  "{\it identical to the cross section in Born Approximation, except for 
the displacement}
\begin{equation}
{\bf q}^2 \to {\bf q}^2_{\rm eff} = \omega^2 + 4\,(E_{\rm i} - {\bar V}_C) 
(E_{\rm f} - {\bar V}_C)\,\sin^2\theta/2\,\,".
\label{effmom}
\end{equation}
By means of a Taylor expansion the Coulomb interaction energy ${\bar V}_C$ is
approximated by $V_C(0)$, i.e. the energy at the center 
of the nucleus ($V_C(0) = -3/2\,Ze^2/R$ for a hard sphere model of the 
nucleus with charge $Ze$ and radius $R$).

The conclusions of  Czy\.z and Gottfried are questionable as well as 
their definition of EMA. In particular one can notice that the displacement 
(\ref{effmom}) implies a modification of the Mott cross section 
\begin{equation}
\sigma_{\rm Mott} = 4\,\alpha^2\,{E_{\rm f}^2 \over q^4}\,\cos^2\theta/2 \to
4\,\alpha^2\,{E_{\rm f}^2 \over q_{\rm eff}^4}\,\cos^2\theta/2\,\,,
\label{eq0}
\end{equation}
as can be seen from their eq.(2.11). This is an artifact originating from 
the form of the eikonal approximation assumed by Czy\.z and Gottfried 
as will be discussed in the next section \ref{sect13}.

\subsection{High-Energy analytical solutions}

\subsubsection{The approach of Yennie, Boos and 
Ravenhall}
\label{sect12}

Yennie, Boos and Ravenhall \cite{YBR65} derived a three-dimensional 
approximation to high-energy electron scattering on nuclei extracting 
an analytical expression valid in the vicinity of the nucleus. The method 
employed an asymptotic expansion in inverse powers of $q R$; the electron 
wave function does not keep the plane wave form and both amplitude and phase 
contain several contributions. In particular current conservation introduces a factor
$k_{\rm eff}/k$ which modifies the electron wave functions at lowest order, an 
effect not considered by Czy\.z and Gottfried and which has deep consequences 
on the modifications of the cross section as I am going to illustrate.

\subsubsection{The synthesis of Rosenfelder and the Mott cross section}
\label{sect13}

Rosenfelder \cite{roland80}, in his comprehensive paper on 
quasielastic electron scattering, discussed also Coulomb corrections
making use of a high-energy electron wave function \cite{lenzrol71} 
based on the formulation due to Yennie \etal. Referring to their own 
work \cite{lenzrol71}, he wrote that "{\it for high-energy electrons 
the distorted wave can be approximated by
\begin{equation}
\psi_{\bf k}({\bf r}) = {|{\bf k}_{\rm eff}| \over |{\bf k}|}\, 
e^{i\,{\bf k}_{\rm eff} \cdot {\bf r}}
\;\;\; {\rm with}\;\; {\bf k}_{\rm eff} = {\bf k} -\hat {\bf k} {\bar V}_{\rm C}\,\,,
\label{eq1}
\end{equation}
where ${\bar V}_{\rm C}$ is a mean value of the electrostatic potential of the nucleus
($\bar V_{\rm C} \approx -3\,Z\alpha/2\,R$ with $R=(5/3)^{1/3}\,\langle r^2\rangle^{1/2}$
for a nucleus with charge $Z$ and rms-radius $\langle r^2\rangle^{1/2}$). The net 
effect is the replacement} ${\bf q} \to {\bf q}_{\rm eff}$
{\it as argument in the structure functions. Note that the amplitude factor 
$|{\bf k}_{\rm eff}| / |{\bf k}|$ makes sure that the Mott cross section remains 
unchanged.} 

The synthesis proposed by Rosenfelder has many practical consequences, namely:

\noindent i) the leading form of the electron wave 
function (\ref{eq1}) incorporates, in addition to the
effective momentum, the change in amplitude due to the 
focusing of the wave front also discussed by Yennie \etal;

\noindent ii) both incoming and outgoing leading focusing corrections
to the electron wave functions
concur to preserve the Mott cross section in its classical form. In fact the 
two terms (${|{\bf k}_{\rm i,f, eff}| / |{\bf k}_{\rm i,f}|}$ of eq.(\ref{eq1})), 
factorize in calculating the ($e,e'$) cross section and are absorbed in the 
Mott expression: an important difference with respect the simplified assumptions of 
Czy\.z and Gottfried (cf. eq.(\ref{eq0})). In detail:
\begin{eqnarray}
d \sigma & = & 4 \alpha^2\, {\cos^2\theta/2 \over q_{\rm eff}^4}\, 
\left({|{\bf k}_{\rm i, eff}| \over |{\bf k}_{\rm i}|}\,
{|{\bf k}_{\rm f, eff}| \over |{\bf k}_{\rm f}|}\right)^2\,
d{\bf k}_{\rm f} \sum_n {1\over2} \sum_{\lambda_i,\lambda_f}
\left|W_{n0}\right|^2\,\delta\left(E_{n0} - \omega \right) = \nonumber \\
& = & 4 \alpha^2 \,{\cos^2\theta/2 \over q^4}\,E_{\rm f}^2\,
d E_{\rm f}\,d\Omega_{\rm f} \sum_n {1\over2} \sum_{\lambda_i,\lambda_f} 
\left|W_{n0}\right|^2\,\delta\left(E_{n0} - \omega \right) = \nonumber \\
& \equiv & \sigma_{\rm Mott}\,d E_{\rm f}\,d\Omega_{\rm f} 
\sum_n {1\over2} \sum_{\lambda_i,\lambda_f} 
\left|W_{n0}\right|^2\,\delta\left(E_{n0} - \omega \right)  \,\,,
\label{smottclassical}
\end{eqnarray}
where
\begin{equation}
W_{n0} = {1 \over \cos\theta/2} \int d{\bf x}\,
e^{i {\bf q}_{\rm eff}\cdot {\bf x}}\,
\bar u_{\lambda_i}({\bf k}_{\rm i,eff}) \gamma_\mu \bar u_{\lambda_f}({\bf k}_{\rm f,eff})\,
\langle n|J^\mu({\bf q}_{\rm eff})|0\rangle
\end{equation}
is the usual matrix element of the transition current\cite{DeFoWa66} for free electrons
with momenta ${\bf k}_{\rm i,f,eff}$.

On the contrary the Effective Momentum Approximation procedure proposed by 
Czy\.z and Gottfried would imply a modification of the Mott cross section 
which must be further compensated by considering the renormalization of the 
incoming and outgoing electron flux before comparing theory with data. The 
emphasis I am giving to this point is not academic; the confusion on that 
specific aspect is at the origin of incorrect experimental analysis as I 
will discuss in section \ref{sect4Jou-Ca};

\noindent iii) Rosenfelder mentions, as interaction energy to be used
in the definition of the effective electron momentum and energy, a
{\it mean} value of the Coulomb potential; a choice which differs 
from the popular assumption of the value at the origin 
$V_{\rm C}(0)$ (a formal mathematical consequence of the expansion of the 
wave function). The practical value he assumes is again the central value 
of an hard sphere model (cf. his discussion after eq.(\ref{eq1})), but the 
intuition is basically correct and I will discuss this aspect again in 
section \ref{sect2emexp}.

\section{DWBA cross section}
\label{sect2}

\subsection{Higher Order effects}
\label{sect2hoeffects}

The contribution we gave to the problem of finding an 
approximate expression for the DWBA cross section is strongly based on the path 
summarized in the previous points. The step forward made is the inclusion of the 
relevant additional focusing terms {\bf beyond} the simple leading factors 
${|{\bf k}_{\rm i,f, eff}| / |{\bf k}_{\rm i,f}|}$ of eq.(\ref{eq1}), terms which 
modify the phase of the electron waves as discussed by Yennie \etal~\cite{YBR65} 
and Lenz and Rosenfelder  \cite{lenzrol71}
(cf. sections \ref{sect12} and \ref{sect13}).
Generalizing a method proposed by Knoll \cite{Knoll} for the investigation of the 
transition form factors to discrete states, the analytic solution of ref.\cite{lenzrol71} 
has been used to expand the DWBA matrix elements in terms of the Born solution and its 
derivative with respect the momentum transfer and applied to exclusive $(e,e',p)$ 
as well as inclusive $(e,e')$ quasielastic scattering \cite{TrTu87,TrTuZg88}. 
The approach has been developed up to second order in $Z\alpha$ and leads to an 
approximated but transparent way of writing the DWBA $(e,e')$ cross section, namely:
\begin{eqnarray}
& &\left. {d\sigma \over d E_{\rm f} d\Omega_{\rm f}}\right|_{\rm DWBA} 
\approx \sigma_{\rm Mott}\,\left\{\left({q_{\rm eff}^2\over {\bf q}_{\rm eff}^2}\right)^2
\,S_L({\bf q}_{\rm eff},\omega)\,\left[1+\Delta_L ({\bf q}_{\rm eff}, 
\omega, E_{\rm i, eff})\right] \, + \right.
\nonumber \\
& + & \left. \left[ -{q_{\rm eff}^2\over 2 {\bf q}_{\rm eff}^2}+
\tan^2 {\theta\over 2}\right]\,S_T({\bf q}_{\rm eff},\omega)\,\left[1+\Delta_T 
({\bf q}_{\rm eff},\omega, E_{\rm i, eff})\right]\,+ 
S_{\rm int}({\bf q}_{\rm eff}, \omega, E_{\rm i, eff}) \right\}\,\,.
\label{dwbacs}
\end{eqnarray}
Equation (\ref{dwbacs}) is, as a matter of fact, close to 
the Rosenfelder's conclusions because $\sigma_{\rm Mott}$  assumes 
its classical form (cf. eq.(\ref{smottclassical})) and the effective momentum 
transfer ${\bf q}_{\rm eff}$ replaces the kinematical momentum ${\bf q}$ as 
argument in the structure functions. However additional modifications appear 
and they are embodied in the terms $\Delta_L$, $\Delta_T$ and in a 
Longitudinal-Transverse interference contribution 
$S_{\rm int}({\bf q}_{\rm eff}, \omega, E_{\rm i, eff})$.  All these terms 
derive from higher order focusing contributions in the high-energy expansion 
of the electron waves and prevent the separability of the DWBA cross section. 
The size of their contribution is crucial to understand the limit of the PWBA 
approximation and the role of the effective momentum transfer. 

A detailed calculation performed in a simple model of quasielastic 
scattering\cite{Tr95}, suggests that
the interference contribution $S_{\rm int}$ is negligible in the whole 
kinematical range of interest also for nuclei as large as $^{208}$Pb
($\lesssim$ 0.01\% with respect to $S_L$ and $S_T$) and also the contributions
$\Delta_L$ and $\Delta_T$ are rather small (remaining within 0.5\% in the
quasielastic peak region and reaching 2 - 4\% for the high-$\omega$ region
and forward angles or for low-$\omega$ and backwards angles).
These small deviations can play some minor role in the longitudinal transverse
separation of the cross section and for a discussion I refer the reader to
the papers of ref.\cite{Tr95}. Of course the estimation of the absolute values
of $\Delta_L$, $\Delta_T$ and $S_{\rm int}$ are model dependent and they can differ
for more sophisticated models of $(e,e')$ reactions. However the relative sizes are
much more independent and the conclusion on their tiny contributions
can be considered reliable.

\subsection{EMA: the result of an asymptotic expansion}
\label{sect2ema}

The marginal role of higher order corrections reduces the cross 
section (\ref{dwbacs}) to a simplified and separable form valid 
(in particular) for medium-weight nuclei:
\begin{equation}
\left.{d\sigma \over d E_{\rm f} d\Omega_{\rm f}}\right|_{\rm DWBA} \approx 
\left.{d\sigma \over d E_{\rm f} d\Omega_{\rm f}}\right|_{\rm EMA} = 
\sigma_{\rm Mott}\left\{\left({q_{\rm eff}^2\over {\bf q}_{\rm eff}^2}\right)^2
S_L({\bf q}_{\rm eff},\omega) + 
\left[ -{q_{\rm eff}^2\over 2 {\bf q}_{\rm eff}^2}+
\tan^2 {\theta\over 2}\right]S_T({\bf q}_{\rm eff},\omega) \right\}.
\\
\label{emacs}
\end{equation}
I will call the approximation (\ref{emacs}) Effective Momentum Approximation 
(EMA) in analogy with my previous works \cite{TrTuZg88,Tr95}. 
However let me stress 
that the explicit form of the electron wave function responsible for the 
reduction (\ref{emacs}) contains also the flux renormalization factors
of eq.(\ref{eq1}) in order to preserve current conservation, a factor 
which also preserves the classical form of the Mott cross section
(cf. eq.(\ref{smottclassical})). 
Another interesting point must be kept in mind: the expansion
which produces the 
analytical result (\ref{dwbacs}) is an asymptotic expansion.
The effective momentum transfer ${\bf q}_{\rm eff}$ has to be 
chosen close to the "real" momentum transfer (which differs from the 
kinematical momentum ${\bf q} = {\bf k}_i - {\bf k}_f$  as measured
in the laboratory) in such a way 
that the transition matrix elements of the nuclear current become 
smooth functions in the neighbourhood of $r=0$ and the coefficients 
of the expansion tend soon to zero \cite{Knoll}. 
Since the effective momentum is a phenomenological quantity, its value
has to be deduced from experimental evidences and eventually
justified, from a theoretical point of view, only \'a posteriori.
That is why most of the authors followed the mathematical 
guide, due to the expansion around $r=0$, by choosing $V_C(0)$ as 
correction terms in the definitions (\ref{effmomv}) and (\ref{eq1}).
The way to know the "real" momentum transfer in quasielastic 
scattering off heavy nuclei is to measure it so that eq.(\ref{emacs}) assumes
all its relevance only after the experimental determination of $\bar
V_{\rm C}$.

\subsection{The effective momentum from experiments}
\label{sect2emexp}

Gu\`eye \etal~reported on a dedicated experiment \cite{gueyethesis} 
performed at the Saclay 
linear accelerator and recently published \cite{gueyeetal99}.
Inclusive quasielastic $(e,e')$ cross sections on $^{12}$C and $^{208}$Pb 
have been measured using electron and positron beams in order to 
investigate charge dependent Coulomb corrections. Gu\`eye \etal~have
been able to measure both the lowest order correction (determining
the Coulomb interaction energy ${\bar V}_{\rm C}$) and higher order
effects. These last contributions turn out to be quite small ($\sim$ 3\%) 
once the effective kinematics is extracted from the data and the
EMA of eq.(\ref{emacs}) used to determine the total response. At the same time
the experiment shows that the Coulomb potential energy
related to the effective kinematics is quite close to the average
\begin{equation}
{\bar V}_{\rm C} = {\int d^3 r\,V_{\rm C}(r)\,\rho_{\rm charge}(r)\over 
\int d^3 r\,\rho_{\rm charge}(r)}
\label{Vaverage}
\end{equation}
an observation which definitely substantiates the {\it mean value} idea 
proposed by Rosenfelder (cf. section \ref{sect13}). In particular,
in the case of for $^{208}$Pb,
$|{\bar V}_{\rm C}| = 18.9\pm 1.5$ MeV from the experiment and 
$|{\bar V}_{\rm C}| = 20.1$ MeV from eq.(\ref{Vaverage}),
(while $|{V}_{\rm C}(0)| = 25.9$ MeV).

\noindent A few comments are in order:

\begin{itemize}

\item[i)] the experiment of Gu\`eye \etal~corroborates the EMA
scheme discussed in the previous sections;

\item[ii)] the information on the "real" value of the momentum transfer
or, equivalently, of the average Coulomb interaction energy, validates
eq.(\ref{emacs}) as the approximated separable form of the $(e,e')$
cross section for medium-weight and heavy nuclei as long as few percent
residual effects (due to higher order focusing effects) can be neglected.

\end{itemize}

\section{Coulomb corrections: the numerical approach}
\label{sect3}

A rigorous treatment of Coulomb distortion 
can be performed by means of a direct numerical calculation of 
the DWBA matrix elements of the nuclear current.

\subsection{The DWBA calculation of Co' and Heisenberg}

The first complete numerical attempt for quasielastic scattering 
is due to Co' and Heisenberg \cite{CoHei87} and they conclude
that the separability of the cross section is definitely lost
in DWBA, in particular for the transverse response. 
A reliable model of the nuclear excitation in the quasielastic 
region is needed in order to extract the correction factors
due to Coulomb distortions. This pessimistic conclusion on the 
model independence of the response functions is due to the 
comparison of their complete calculation with the PWBA results and to 
the idea that focusing contributions, added by the DWBA description, cannot
be disentangled with the necessary precision in a model independent way. 
In fact Co' and Heisenberg assume that the eikonal approximation 
{\it \`a la} of Czy\.z and Gottfried (cf. section \ref{sect11}) is the 
only analytic approximation one can make. The relevance of the flux
renormalization effects, included in the DWBA and not in the 
eikonal approach, prevents the definition of an effective momentum 
approximation while the nature of the numerical solution cannot manifest
the structure of cross section in DWBA.

However the contribution of Co' and Heisenberg remains fundamental. 
The conclusion that the Rosenbluth plot of the DWBA appears to
be linear despite the non separability of the cross section
is illuminating.  At that time such information was known 
\cite{TrTu87}, but within an approach including terms up to 
$(Z\alpha)^2$ only and not for a complete DWBA 
calculation. The fact that the usual Rosenbluth plot of a complete 
DWBA calculation shows a rather close 
linearity \cite{CoHei87,TrTu87,TrTuZg88} 
demonstrates that experimental evidence of linearity is not a sufficient 
condition for the separation of the cross section. A large effect of the 
Coulomb distortion on the Rosenbluth representation is a rotation of the 
straight line whose intercept and slope are no longer connected with the 
longitudinal and transverse responses.

\subsection{1996: the approach of Kim \etal}
\label{sect3kimetal}

Also Kim \etal~\cite{ohio96} have the advantage 
of an exact DWBA numerical solution, but they have a more 
ambitious project: extracting an approximated form of the
cross section which is not as time consuming as the complete
DWBA procedure but able to reproduce the exact results keeping, 
at the same time, a separable form. 
The main ingredient is the local momentum transfer. In fact
the momentum of the electron, in the external Coulomb field,
can be rigorously defined as a local ($r$-dependent) quantity
only. Consequently a Local Effective Momentum Approximation
(LEMA) is considered in ref.\cite{ohio96} as leading approximation 
instead of the
EMA which involves an average over the nuclear volume.
To transform this scheme into a simple form of DWBA cross 
section the authors need a certain number of {\it ad hoc} 
assumptions such as: 

\begin{itemize}

\item[i)] the cross section is {\it \'a priori} assumed 
to have a separable form of Rosenbluth type. Interference
contributions like those included in eq.(\ref{dwbacs}), 
are simply not considered\footnote{The analytic approach of ref.\cite{Tr95}
				shows that the specific error is, in practice, 
				irrelevant because the 
				interference contributions are 
				generally negligible. However this
				conclusion cannot be drawn from a
				numerical calculation and is, in any case,
				valid {\it \'a posteriori} only.}. 
Moreover the structure functions, which depend, to a 
good approximation, on the effective momentum and energy 
transfer only, (cf. eq.(\ref{emacs})) are charged of an artificial
dependence on the kinematical conditions (incident energies 
and angles) due to the choice of factorizing the Rosenbluth
terms 
$\left({q^2\over {\bf q}^2}\right)^2$ and
$\left[ -{q^2\over 2 {\bf q}^2}+\tan^2 {\theta\over 2}
\right]$
instead of the averaged quantities
$\left({q_{\rm eff}^2\over {\bf q}_{\rm eff}^2}\right)^2$
and
$\left[ -{q_{\rm eff}^2\over 2 {\bf q}_{\rm eff}^2}+
\tan^2 {\theta\over 2}\right]$;

\item[ii)] the local effective momentum transfer is assumed 
along the kinematical momentum transfer, a choice valid for 
elastic scattering only. More precisely, the inelastic effective momentum
of eq.(\ref{effmomv}) can be written
\begin{equation}
{\bf q}_{\rm eff} = {\bf q} +\left(\hat {\bf k}_i - \hat {\bf k}_f\right)
{\bar V}_{\rm C} = {\bf q} \left(1 - {\bar V_{\rm C}\over E_i}\right)
+ {\omega \over E_i} {\bar V}_{\rm C} \hat {\bf k}_f\,\,,
\label{effmomine}
\end{equation}
and becomes ${\bf q}_{\rm eff} = {\bf q} \left(1 - {\bar V_{\rm C}\over E_i}\right)$
only for $\omega = 0$.

The assumption of Kim \etal~introduces deviations comparable with higher order
focusing corrections (in particular at backward angles). 
As an example, for $\theta = 140^0$ and momentum transfer as high
as 400 MeV/c, the cross section is enhanced
by $\approx 4$\% on top of the quasielastic peak and 
reduced by $\approx 20$\% at higher energy transfer
(when its value reaches 1/3 of the maximum);

\item[iii)] several {\it ad hoc} factors
are introduced in the approximated LEMA expression
to reproduce the DWBA results. Most of them are tuned 
on the DWBA electron cross sections and they 
can induce quite different effects in the case of positron 
scattering partially explaining the discrepancy 
of the LEMA calculation by Kim \etal~with
the experimental results of ref.\cite{gueyeetal99}.

\end{itemize}

\subsection{{\it faiblesse} of the numerical approach}

The merit of a complete numerical approach is obvious;
nevertheless there is a {\it point de faiblesse},
already visible in the paper by Co' and
Heisenberg. It is determined by the intrinsic difficulty 
in separating higher order (focusing) effects from the 
simple flux renormalization due to the factors
$|{\bf k}_{\rm i,f, eff}| / |{\bf k}_{\rm i,f}|$ 
of eq.(\ref{eq1}) (both the effects are of course included in 
the complete solution). The natural leading reduction of a full
DWBA numerical calculation appears to be the eikonal 
approximation of Czy\.z and Gottfried 
(cf. section \ref{sect11}) with the consequence of a
change of the Mott cross section and the need of a
renormalization of incoming and outgoing flux. This is a 
complicated procedure which can become source of errors
in the analysis of the experimental data, as I will
show in the next section.

\section{Analysis of the experimental data}
\label{sect4}

Some of the experimental data on quasielastic electron scattering for
medium weight and heavy nuclei have been analyzed including Coulomb 
distortion effects. In particular the Bates experiment on $^{238}$U\cite{bla86},
the Saclay data on $^{208}$Pb\cite{ZgetalPb}, the reanalysis
of Jourdan\cite{jourdan}, and the $^{40}$Ca experiment at 
Bates\cite{batesca40}.
In this section I will summarize the situation
to ask for new analysis which include Coulomb effects in a more 
consistent and/or more reliable way.

\subsection{Bates data on $^{238}$U}
\label{sect4U}

The first attempt of obtaining quasielastic ($e,e'$) data on an heavy 
nucleus dates back to an experiment performed at Bates on $^{238}$U\cite{bla86}. 
The data have been analyed by means of an effective momentum transfer.
However the approximation adopted was just a generalization of the scheme known 
for elastic scattering and the effective momentum was chosen to be along the 
kinematical momentum transfer, a choice valid for elastic scattering only as 
discussed already in section \ref{sect3kimetal} (cf. eq.(\ref{effmomine})).

Also focusing corrections were included by adapting a phase shift code used 
for elastic scattering. The details of the procedure are discussed neither 
in the article nor in the PhD thesis of Blatchley. The approach, however, has 
the merit of a first attempt even if manifestly insufficient for a complete analysis.

\subsection{Saclay data on $^{208}$Pb}
\label{sect4Pb}

The Saclay data on $^{208}$Pb \cite{ZgetalPb} have been analyzed
including the quasielastic effective momentum and higher order corrections 
systematically. In particular the EMA in the form given by eq.(\ref{emacs}) 
is used, for the first time, as leading order approximation to disentangle 
Coulomb corrections. Higher order effects are also discussed and included 
within the approximations of ref.\cite{TrTuZg88}. These approximations are, 
however, too severe to give a quantitative account of higher order contributions 
and a more rigorous treatment of the transition matrix elements shows\cite{Tr95} 
that the effects are smaller.
Also in the case of $^{208}$Pb a reanalysis is, therefore, useful even if the 
expected modifications cannot change the qualitative conclusions drawn in 1994.

\subsection{Jourdan's data analysis and the Bates experiment on $^{40}$Ca}
\label{sect4Jou-Ca}

I discuss the two analysis together for two
reasons: i) they refer to medium-weight nuclei; ii) they both make use
of the Coulomb distortion analysis proposed by Kim \etal~, i.e. the use of
a numerical approach. In the first case
the method and the corrections suggested by Kim \etal~ have been applied to 
existing data on $^{12}$C \cite{C12}, $^{40}$Ca \cite{Ca40,CaFe} and 
$^{56}$Fe \cite{CaFe,Fe56}, while in the latter Jin, Wright 
and Onley coauthored the paper on the experiment. 

Both papers discuss first the problem of flux 
renormalization due to leading order focusing effects (i.e. 
the factors ${|{\bf k}_{\rm i,f, eff}| / |{\bf k}_{\rm i,f}|}$ 
I discussed in section \ref{sect13}). The complications induced
by a numerical approach appear immediately: the correction factor
chosen\footnote{
				More precisely the two papers 
				describe the application of two 
				opposite procedures: the
				data of the ($e,e'$) cross section
				are {\bf reduced} by a factor
				$({|{\bf k}_{\rm i}| / |{\bf k}_{\rm i, eff}|})^2$
				in the Jourdan analysis and
				{\bf increased} by the same
				quantity in the Bates paper: however 
				a closer look at the data analysis 
				supports the idea of a misprint 
				in the paper on $^{40}$Ca\cite{MorgenPr}.}
to renormalize the cross section data is
$({|{\bf k}_{\rm i}| / |{\bf k}_{\rm i, eff}|})^2$ a choice which 
involves the incident energies only and it has
no theoretical justification. In fact if the EMA
of eq.(\ref{emacs}) is accepted as leading approximation, the
Mott cross section should be kept unchanged and higher order effects
estimated. On the contrary if one prefers to obtain
the cross section in the eikonal approximation of Czy\.z and 
Gottfried because this is the leading part of the numerical calculation, 
both the incoming and outgoing electron waves must be 
renormalized and not the incoming flux only (cf. section \ref{sect13}). 
Actually a possible origin of the incorrect normalization procedure is in
the way of writing the Mott cross section in both papers, namely
$\left({\alpha \cos\theta/2 \over 2 E_{\rm i}\sin^2\theta/2}\right)^2$.
The correct expression which distinguishes the contributions of two
different regions (the interaction volume and the detector) reads
$\left(2 {\alpha E_{\rm f} \cos\theta/2 \over q^2}\right)^2$ and 
it reduces to the previous form only if Coulomb corrections are negligible.
In fact the term $E_{\rm f}^2$ in the numerator of eq.(\ref{eq0}) originates
from the detection volume $d{\bf k}_{\rm f}=E_{\rm f}^2\,d \Omega_{\rm f}\,
d E_{\rm f}$ in the Lab, while the propagator of the virtual photon,
$\sim {1\over q^2}$, involves the interaction region where the motion of 
the electron is influenced by the Coulomb potential and is to be modified
to $\sim {1\over q_{\rm eff}^2}$ as stated by Czy\.z and Gottfried
(see discussion in section \ref{sect11}). The inclusion of the leading 
focusing terms of eq.(\ref{eq1}) compensate such a modification as already 
discussed (cf. eq.(\ref{smottclassical})).

The manifest inconsistent treatment of Coulomb distortions
influences the conclusions of the analysis of refs.\cite{jourdan,batesca40};
a reanalysis would be welcome.

\section{Final remarks and conclusion}
\label{concl}

The structure of the DWBA cross section can be reduced (up to order 
$(Z\alpha)^2$) to the form (\ref{dwbacs}). The contributions $\Delta_{L,T}$ 
and $S_{\rm int}$ are due to higher order focusing effects of 
the electron waves in 
the proximity of the nucleus, in particular to its phase deformation. 
On the contrary the renormalization of the electron waves due to current
conservation is a leading order effect and can be incorporated in a simple 
form (the expression (\ref{emacs})) known as Effective Momentum Approximation 
(EMA) where only higher order effects are neglected. The fact that in 
eq.(\ref{emacs})
the Mott cross section keeps its classical expression is just a byproduct
of current conservation.
The Effective Momentum Approximation is a good scheme to interpret inclusive
data as experimentally verified in the recent analysis of
electron and positron quasielastic scattering\cite{gueyeetal99} and theoretically
predicted in a series of papers\cite{TrTuZg88,Tr95}.

Quasielastic data should be reanalyzed within such scheme in 
a consistent way to include those Coulomb distortion effects which give sizable
contributions in the separation of the cross section in longitudinal 
and transverse components. The recent application\cite{jourdan,batesca40} 
of more complete numerical DWBA results\cite{ohio96} shows 
a clear inconsistency and the data on longitudinal/transverse structure functions
cannot be considered reliable, a remark confirmed by the comparison of the
calculation by Kim \etal~with the experimental data on Coulomb corrections
measured comparing quasielastic scattering by electrons and positrons off
$^{12}$C and $^{208}$Pb\cite{gueyeetal99}.

\section*{Acknowledgments} Interesting discussions with J. Morgenstern and 
a useful correspondence with C. Williamson and J. Jourdan are gratefully
acknowledged. I thank G. Orlandini and V. Vento for a critical 
reading of the manuscript.

\end{document}